\documentclass[aps,prl,twocolumn,showpacs,preprintnumbers,amsmath,amssymb,graphicx,
superscriptaddress]{revtex4}
\usepackage{epsfig}
\usepackage{dcolumn}% Align table columns on decimal point
\usepackage{bm}% bold math
\usepackage{natbib}
\usepackage{graphicx}
\usepackage{subfigure}
\usepackage{sidecap}

\begin{document}

\author{G.\ Karapetrov}\email{goran@anl.gov}
\affiliation{Materials Science Division, Argonne National
Laboratory, Argonne, Illinois 60439, USA}
\author{M. V. Milo\v{s}evi\'{c}}
\affiliation{Departement Fysica, Universiteit Antwerpen,
Groenenborgerlaan 171, B-2020 Antwerpen, Belgium}
\author{M.\ Iavarone}
\affiliation{Materials Science Division, Argonne National
Laboratory, Argonne, Illinois 60439, USA}
\author{J.\ Fedor}
\affiliation{Materials Science Division, Argonne National
Laboratory, Argonne, Illinois 60439, USA}\affiliation{Institute of
Electrical Engineering, Slovak Academy of Sciences, Dubravska cesta
9, 841 04 Bratislava, Slovakia}
\author{A.\ Belkin}
\affiliation{Materials Science Division, Argonne National
Laboratory, Argonne, Illinois 60439, USA} \affiliation{Physics
Division, Illinois Institute of Technology, Chicago, Illinois 60616,
USA}
\author{V.\ Novosad}
\affiliation{Materials Science Division, Argonne National
Laboratory, Argonne, Illinois 60439, USA}
\author{F. M. Peeters}
\affiliation{Departement Fysica, Universiteit Antwerpen,
Groenenborgerlaan 171, B-2020 Antwerpen, Belgium}

\title{Transverse instabilities of multiple vortex chains\\ in superconductor-ferromagnet bilayers}
\date{\today}
\tighten

\begin{abstract}
Using scanning tunneling microscopy and Ginzburg-Landau simulations,
we explore vortex configurations in magnetically coupled
NbSe$_2$-Permalloy superconductor-ferromagnet bilayer. The Permalloy
film with stripe domain structure induces periodic local magnetic
induction in the superconductor, creating a series of
pinning-antipinning channels for externally added magnetic flux
quanta. Such laterally confined Abrikosov vortices form {\it
quasi-1D arrays} (chains). The transitions between multichain states
occur through propagation of kinks at the intermediate fields. At
high fields we show that the system becomes {\it non-linear} due to
a change in both the number of vortices and the confining potential.
The longitudinal instabilities of the resulting vortex structures
lead to vortices {\it `levitating' in the anti-pinning channels}.
\end{abstract}

\date{\today}
 \maketitle

Recently, much attention has been devoted to the studies of hybrid
systems comprising of two or more elements with complementary
physical properties, with the motivation that the resulting hybrid
has a superior performance as compared to its constituents.
Superconductor - ferromagnet hybrids exploit the interaction between
the two `reservoirs' of strongly correlated electrons resulting in a
wealth of new physical phenomena (for review see \cite{buzdin}).
%  ,lyuksyutov,volkov,izyumov,schrev
The fundamental premise has been to isolate and tailor the dominant
interaction between two correlated systems, as many overlapping
interactions may conceal the discovery of new physical phenomena.
Recent explorations focused solely on {\em magnetic interaction}
between superconductor and ferromagnet have led to the discovery of
numerous intriguing phenomena associated with magnetic pinning of
Abrikosov vortices as well as with mesoscopic confinement of
superconductivity~\cite{vvmnatmat, bulaevskiAPL, gillijns,
stamopoulos, milosevic, vlasko_hybrid, belkin_APL}.

Confinement of superconductivity to quasi-1D translates into
confinement of the superconducting vortices. Abrikosov vortex
structures constrained to one-dimensional superconducting condensate
has been actively studied \cite{guimpel, ivlev, kes_channels,
kokubo, field, grigorieva, karapetrov_PRL} in order to examine novel
static and dynamic vortex phases. Magnetic imaging of the vortex
structures in these systems have been challenging, in particular in
the cases where effective penetration depth of the superconductor
(i.e. magnetic signature of the Abrikosov vortex) is on the order of
the intervortex spacing. On the other hand, scanning tunneling
microscopy (STM) can be successfully used to image the distribution
of the \emph{local order parameter} on the surface and map the
Abrikosov vortex distribution in higher magnetic fields.

\begin{figure}
\includegraphics[width=3.0in]{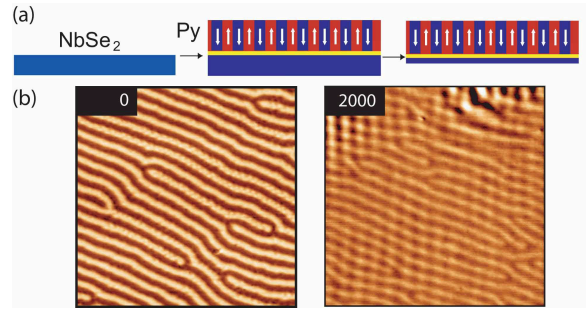}
\caption{(Color online) (a) Fabrication of magnetically coupled
superconductor-ferromagnet structure: on a single crystal NbSe$_2$ a
thin layer of insulating SiO$_2$ is evaporated followed by a 1$\mu
m$ thick Permalloy. After cleaving the NbSe$_2$ down to submicron
thickness, fresh atomically flat surface is exposed to the STM
probe. (b) Magnetic force microscope images of the stripe domain
pattern in Py at 0 Oe and 2000 Oe applied field perpendicular to the
surface of the film at room temperature. Scan area is 20$\times$20
$\mu$m$^2$.} \label{Figure1}
\end{figure}

In this paper, motivated by the aforementioned results, we study the
vortex configurations in magnetically coupled superconductor
(NbSe$_2$)~/ferromagnet (Permalloy) bilayer. We take advantage of
well-ordered magnetic state of the Permalloy (Py) film to induce 1D
vortex confinement in the adjacent superconductor. We combine STM
and Ginzburg-Landau simulations to explore the magnetic interaction
between the magnetic domain state in a ferromagnet and Abrikosov
vortices in an extreme type-II superconductor. NbSe$_2$ was chosen
for its small coherence length (compared to the ferromagnetic domain
size), negligible intrinsic pinning, and atomically flat and inert
surface. The choice of thick Py film with well-ordered stripe domain
pattern allows us to obtain a 1D periodic potential modulation of
the superconducting condensate. Earlier we have shown that such
magnetic potential emanating from the ferromagnet can be used to
achieve domain wall superconductivity \cite{belkin_APL}. Deeper in
the superconducting state, strong interaction of the Abrikosov
vortex lattice with the periodic magnetic domain structure leads to
commensurability effects~\cite{belkin_PRB}.

The bilayer for present study was prepared using the method
elaborated in Ref.~\cite{karapetrov_APL05}. 20 nm thick SiO$_2$
insulating thin film was sputtered using RF magnetron on a freshly
cleaved high quality NbSe$_2$ single crystals (T$_c$=7.2K, RRR=40).
This layer was covered with a $D=1~\mu$m thick Py
(Fe$_{20}$Ni$_{80}$) overlayer using DC magnetron sputtering from a
single target (Fig.~\ref{Figure1}a). As-prepared sample was attached
to a conducting substrate with the ferromagnetic layer down. The
surface of NbSe$_2$ single crystal was cleaved successively until
the color of the surface started changing. This is a first
indication that the thickness of the NbSe$_2$ top layer is within
the wavelength of the visible light. By further cleaving the sample
the areas of different color could be obtained that signify
different thickness of the superconductor~\cite{novoselov_PNAS}. The
final thickness used in the STM studies in this work was c.a. 0.4~\
$\mu$m and this value was verified by cross-sectional cut of the
sample using focused ion beam etching and scanning electron
microscopy metrology.

The thick Py film forms narrow magnetic stripe domain pattern with a
period comparable to the film thickness~\cite{saito, spain}
(Fig.~\ref{Figure1}b). The period of the domain structure $W$ is
proportional to the thickness $D$ of the Py film. The magnetic
stripe domain structure is insensitive to magnetic field applied
perpendicular to the film's surface up to 400 Oe, but forms
checkerboard domains in higher fields (Fig.~\ref{Figure1}b). In our
theoretical calculations, we used the dependence
$W\approx0.0109\sqrt{D}$ cm, following from Ref.~\cite{szym} and
estimated magnetization of the film $M=1620$ G, exchange constant
$A=2.665\cdot10^{-6}$ erg/cm and uniaxial anisotropy constant
$K=2.63\cdot10^5$ erg/cm$^3$.

The theoretical simulations of the described structure were
performed within Ginzburg-Landau (GL) theory. We solve
self-consistently a set of mean field differential equations for the
order parameter $\psi$ and the vector potential ${\bf A}$:
\begin{eqnarray}
&& (-i\nabla-{\bf A})^2\psi = (1-T-|\psi|^2)\psi, \label{gl1}\\
&& -\kappa^2\nabla\times\nabla\times {\bf A} = {\bf j}~. \label{gl2}
\end{eqnarray}
The latter is the Maxwell-Amp\`{e}re equation with a current density
${\bf j}=\Im(\psi^*\nabla\psi)-|\psi|^2{\bf A}$. GL constant
$\kappa$ equals the ratio between the magnetic field penetration
depth $\lambda$ and coherence length $\xi$, and is in general very
large for NbSe$_2$ samples. In above expressions, all distances are
expressed in units of $\xi(T=0)$, the vector potential in
$\phi_0\big/2\pi\xi(0)$, and the order parameter in $\sqrt{-\alpha
/\beta}$ with $\alpha $, $\beta $ being the GL coefficients. For
more details on the numerics and implementation of periodic boundary
conditions we refer to Ref.~\cite{miloPRL}. Note that virial theorem
requires integer number of flux quanta in the simulation region with
periodic boundary, and we accommodate that by fixing the width of
simulation region to the period of magnetic domain structure $W$,
and the length is fitted (at $\approx2W$) according to the applied
magnetic field.

\begin{figure}[b]
\includegraphics[width=3.5in]{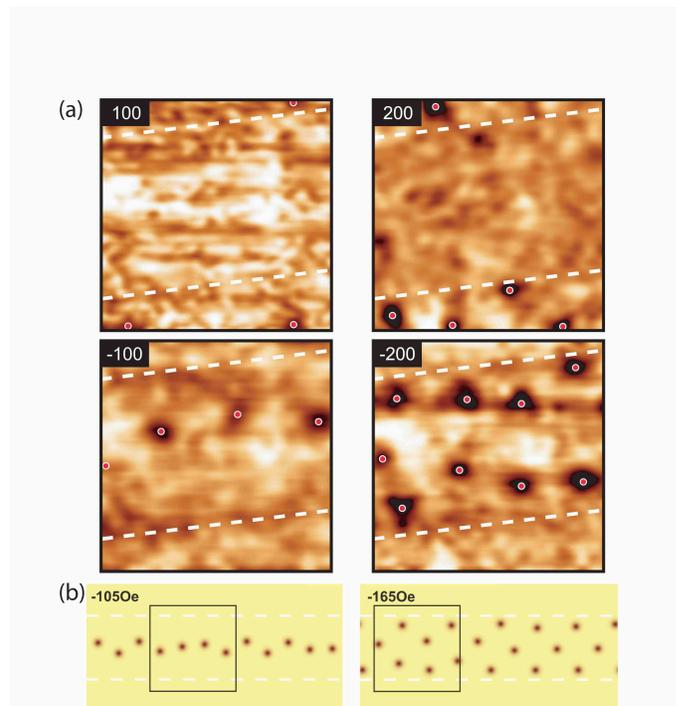}
\caption{(Color online) (a) STM images of vortex configurations in
the NbSe$_2$ at 4.2K. Applied magnetic field values (in Oe)
perpendicular to the surface of the superconductor are shown in the
upper right corner. The white dotted lines show the underlying
magnetic stripe domain boundaries. The scanning area is
$0.7\times0.7~\mu$m$^2$. (b) Cooper-pair density plots, calculated
at -105 and -165 Oe (in comparison with (a), see marked area),
obtained for $D=1~\mu$m (plot size $1.087\times2.176~\mu$m$^2$  and
$1.087\times2.193~\mu$m$^2$, respectively).} \label{Figure3}
\end{figure}

In STM measurements, the vortex images were obtained by spatially
mapping the tunneling conductance value at the edge of
superconducting energy gap at 4.2 K \cite{karapetrov_PRL}. %The
%vortex cores have quasiparticle states within the superconducting
%energy gap and thus lower local density of states peak at the
%superconducting gap edge and higher zero bias conductance. We
%performed a conductance map at $V_{bias} = +2.0$ meV which resulted
%in superconducting areas having higher conductance than the gapless
%regions (dark areas).
In Fig.~\ref{Figure3}, we show the local
density of states (LDOS) map of $0.7\times0.7~\mu$m$^2$ area on
NbSe$_2$ surface. Fig.~\ref{Figure3} (a) shows a slightly distorted
vortex chain at -100 Oe. The effective magnetic field above each
domain is a sum of the contributions of the applied magnetic field
and the field due to the local magnetization of the Py. In the case
when these two contributions are of the same sign the effective
fields are enhanced, and when they are of opposite sign the two
contributions cancel each other. From our map it is obvious that the
central part of the image is above a domain that has a ``negative''
polarity, i.e. the domain is collinear with the negative applied
magnetic field. One expects that at 100 Oe the average vortex
distance should be
$a_v=\sqrt{\left(\frac{2}{\sqrt{3}}\right)}\sqrt{\left(\frac{\Phi_0}{B}\right)}=0.489~\mu
$m, and instead, at -100 Oe the vortices form a quasi-1D chain state
with average distance of $\approx 0.210~\mu$m. No vortices are
observed at +100 Oe in the same area, being pinned at the adjacent
domain.

Further increase of the negative applied field leads to increase in
vortex density and formation of chain states similar to ones
predicted in Ref. \cite{ivlev} and observed earlier in patterned
superconducting NbSe$_2$ \cite{karapetrov_PRL}. The chains consist
of periodically spaced vortices in the middle of the stripe domain.
As magnetic field increases so does the vortex density, leading to
smaller inter-vortex spacing within a single chain. This process is
continuous until a critical point leading to geometrical ordering
transition and formation of double chains (Fig.~\ref{Figure3}(a) at
-200 Oe) and triple chains (Fig.~\ref{Figure5}(a) at -300 Oe). The
similarities in vortex ordering and geometrical phase transitions
with the case of strong 1D confinement potential~\cite{ivlev,
karapetrov_PRL} end at this point. Further, with the help of
simulations, we focus on the subtle but important effects that are
specific to the nature of our confining potential.

\begin{figure}
\includegraphics[width=3.5in]{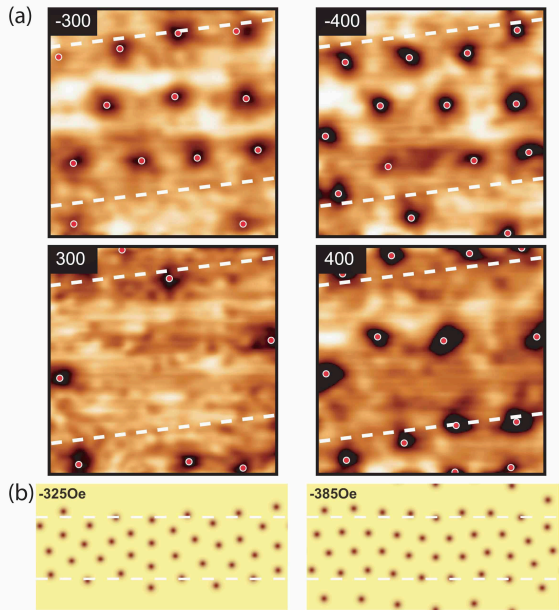}
\caption{(Color online) (a) Same as Fig. \ref{Figure3}, but for
$H=\pm 300$ Oe and $\pm 400$ Oe. (b) Cooper-pair density plots,
calculated at -325 and -385 Oe, for $D=1~\mu$m (plot size
$1.087\times2.176~\mu$m$^2$  and $1.087\times2.193~\mu$m$^2$,
respectively).} \label{Figure5}
\end{figure}

\begin{figure}
\includegraphics[width=3.0in]{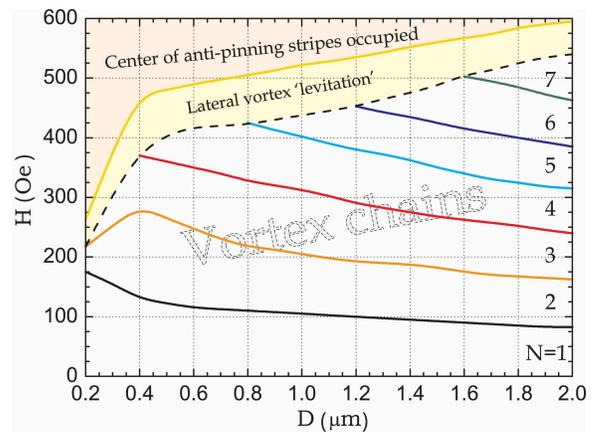}
\caption{(Color online) The equilibrium vortex phase diagram of the
NbSe$_2$ film at $T=4.2$ K, as a function of thickness of the
underlying Py film $D$ and external perpendicular field $H$ (the
domain structure of the Py film is assumed to be unperturbed by the
applied field). $N$ denotes the number of vortex chains along the
positive magnetic domains (illustrated by $|\psi|^2$ contourplots on
the right, dark/light color - low/high density). At high fields,
vortices penetrate areas above negative magnetic domains.}
\label{Figure2}
\end{figure}

We take advantage of the numerical simulations to reconstruct the
vortex phase diagram of the magnetically coupled S/F bilayer system.
Numerical solution of Eqs. (\ref{gl1}-\ref{gl2}) minimizes the Gibbs
free energy of the sample, and Fig.~\ref{Figure2} shows the main
outline of the ground-state phase diagram as a function of thickness
of the Py film. Since this is a 2D calculation, the stray field of
the Py film has been averaged over given thickness of NbSe$_2$ film,
and as such put into Eqs. (\ref{gl1}-\ref{gl2}). The simulation
region was $2W\times W$ large, and simulations were performed for
$\xi(0)=10$ nm and $T=4.2$ K$\approx0.6T_c$. The vortex states obey
the imposed linear confinement and form {\it multiple vortex chains}
($N$ of them) along the stripe domain with out-of-plane
magnetization {\it parallel} to the applied field (as expected from
the theory of magnetic pinning of vortices, see e.g.
Ref.~\cite{miloPRB}). However, it should be emphasized that in
reality transitions between different $N$-chain states are not of
first order. In increasing field, added vortex causes {\it local}
chain instability, by changing locally the transverse ``optical''
mode \cite{liu}. More instabilities eventually lead to a new ordered
$N$-chain state via a so-called "zig-zag" type transition
characteristic for quasi-1D Wigner crystals~\cite{piac}. These
transverse instabilities leading to different ordering are evident
in Fig. \ref{Figure3}, where calculated vortex structures for the
case of $D=1~\mu$m are given for $N=1\rightarrow2$ and
$N=2\rightarrow3$ chain rearrangement, in comparison with
experimentally obtained snapshots of vortex configurations.

Note however that vortex spacing $\Delta x$ along the chain prior to
the $N=2\rightarrow3$ transition remains {\it larger} than
$W/2(N+1)$, a value predicted for $N\rightarrow N+1$ transition in
Ref.~\cite{ivlev}. In Fig. \ref{Figure4} we show the calculated
vortex spacing along the chain prior to the formation of a new chain
as a function of the thickness of the magnetic film (i.e.
proportional to the domain width). The main conclusions following
from Fig. \ref{Figure4} are that (i) threshold vortex spacing
decreases with thickness of the magnetic film, but saturates for
$D>1~\mu$m, and (ii) $\Delta x$ is very weakly dependent of the
number of chains $N$, contrary to findings of Ref.~\cite{ivlev}.
Note that in quasi-1D confined classical systems, with hard-wall
lateral boundaries, $\Delta x$ also decreases with $N$ (see e.g.
\cite{hagh}), while in our system lateral confinement is of
parabolic type (where decrease of $\Delta x$ with $N$ is slower
\cite{piac}).

\begin{figure}[b]
\includegraphics[width=3.5in]{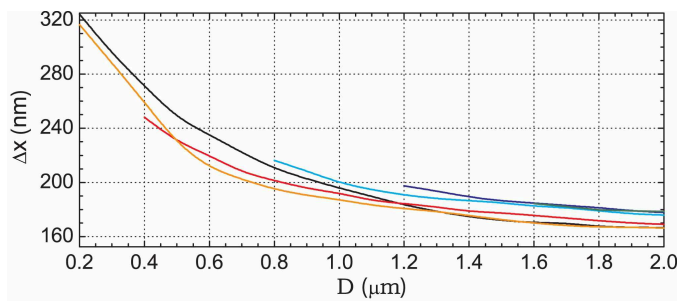}
\caption{(Color online) Average vortex spacing along the chain,
prior to $N\rightarrow N+1$ chains configurational transition (color
coding corresponding to Fig. \ref{Figure2}), as a function of the
thickness of the Py film.} \label{Figure5}
\end{figure}

The discrepancy from the model~\cite{ivlev} is caused by the
non-rigid lateral confinement of the vortex chains. Namely, the
superposition of applied magnetic field and the stray field of the
magnetic domains leads to slow expansion of the confinement
potential for the superconducting vortices. From the magnetic force
microscopy images at room temperature~\cite{belkin_APL} and the STM
maps in Fig.~\ref{Figure3} we can reconstruct the position of the
stripe domains under the NbSe$_2$ layer, and the domain boundaries
are outlined with dashed lines in Figs.~\ref{Figure3}-\ref{Figure2}.
However, beyond $N=3$ one notices that vortex chains may already
occupy areas slightly beyond the domain boundaries. Therefore, with
increasing magnetic field, not only that new vortices are added to
the system, but confinement potential is broadened as well. This
leads to further lateral instability of the chains, in the areas of
strongest interaction with effective confinement. Moreover, the
attractive force acting on newly added vortices may gradually become
{\it comparable} to the overall long-range interaction with earlier
formed vortex chains, forcing the new vortices to reside in the {\it
anti-pinning} channel above the magnetic domain of opposite
polarity. This is shown in Fig.~\ref{Figure5}, for fields $\pm300$
Oe in experiment and 325 Oe in the simulation. Due to competing
interactions, several vortices are `levitating' above the
energetically non-favorable domain. Beyond the field of 325 Oe in
the simulations, we were unable to recover the linear chain state,
unless we had the commensurate number of vortices in each chain. In
every other case, the attempt of the large number of vortices to
form an Abrikosov lattice in competition with linear confinement,
resulted in a sinusoidal-like modulation of the vortex structure,
such as one shown in Fig. \ref{Figure5}(b).

Further increase of the applied field leads to a larger number of
vortices in the chains as well as above the anti-pinning domain. At
certain conditions vortices at the anti-pinning channel may form a
1D structure, being equally repelled (on average) by chain
structures from both adjacent domains. This is a very peculiar
state, since vortex {\it repulsion} in the anti-pinning domain is
{\it strongest} in its central region. The existence of such vortex
ordering is verified in both theory (at $\approx 385$ Oe) and
experiment (at 400 Oe), as shown in Fig.~\ref{Figure5}. Beyond this
field, vortices have little difficulty trespassing to the
anti-pinning domain, which finally leads to a somewhat distorted
Abrikosov lattice across the whole film.

To summarize, we studied the Abrikosov vortex configurations in
magnetically coupled superconductor-ferromagnet bilayers.  We
elucidate the mechanism of quasi-1D topological vortex phase
transitions in this system by theoretical simulations and STM vortex
imaging. We demonstrate that transitions between 1D ordered vortex
phases are accompanied by local transverse instabilities resulting
in local vortex disorder. This, in combination with magnetic
field-relaxed confinement, leads to lower threshold vortex densities
for the geometrical transitions than in strongly confined systems
such as superconducting stripes and hard-wall classical clusters. At
higher magnetic fields the change of number of vortices is
accompanied with the change in confinement potential resulting in
nonlinear system response and vortex chain formation above the
domains with opposite polarity. As a result, we found the
`levitation' of vortices above the magnetic domains of opposite
polarity, as well as vortex chain formation in this energetically
unfavorable region, while the earlier formed chains are being
sinusoidally distorted from the equilibrium Abrikosov vortex
lattice.

This work as well as the use of the Center for Nanoscale Materials
and the Electron Microscopy Center at Argonne National Laboratory
were supported by UChicago Argonne, LLC, Operator of Argonne
National Laboratory (``Argonne''). Argonne, a U.S. Department of
Energy Office of Science laboratory, is operated under Contract No.
DE-AC02-06CH11357. M.V.M. and F.M.P. acknowledge support from the
Flemish Science Foundation (FWO-Vl), the Belgian Science Policy, the
JSPS/ESF-NES program, the ESF-AQDJJ network, and the Vlaanderen-USA
bilateral program.

\bibliographystyle{apsrev}

\begin{thebibliography}{32}
\expandafter\ifx\csname
natexlab\endcsname\relax\def\natexlab#1{#1}\fi
\expandafter\ifx\csname bibnamefont\endcsname\relax
  \def\bibnamefont#1{#1}\fi
\expandafter\ifx\csname bibfnamefont\endcsname\relax
  \def\bibfnamefont#1{#1}\fi
\expandafter\ifx\csname citenamefont\endcsname\relax
  \def\citenamefont#1{#1}\fi
\expandafter\ifx\csname url\endcsname\relax
  \def\url#1{\texttt{#1}}\fi
\expandafter\ifx\csname urlprefix\endcsname\relax\def\urlprefix{URL
}\fi \providecommand{\bibinfo}[2]{#2}
\providecommand{\eprint}[2][]{\url{#2}}

\bibitem[{\citenamefont{Buzdin}(2005)}]{buzdin}
\bibinfo{author}{\bibfnamefont{A.~I.} \bibnamefont{Buzdin}},
  \bibinfo{journal}{Rev. Mod. Phys.} \textbf{\bibinfo{volume}{77}},
  \bibinfo{pages}{935} (\bibinfo{year}{2005});
%\bibitem[{\citenamefont{Lyuksyutov and Pokrovsky}(2005)}]{lyuksyutov}
\bibinfo{author}{\bibfnamefont{I.~F.} \bibnamefont{Lyuksyutov}}
  \bibnamefont{and} \bibinfo{author}{\bibfnamefont{V.~L.}
  \bibnamefont{Pokrovsky}}, \bibinfo{journal}{Adv. Phys.}
  \textbf{\bibinfo{volume}{54}}, \bibinfo{pages}{67} (\bibinfo{year}{2005});
%\bibitem[{\citenamefont{Bergeret et~al.}(2005)\citenamefont{Bergeret, Volkov,
%  and Efetov}}]{volkov}
\bibinfo{author}{\bibfnamefont{F.~S.} \bibnamefont{Bergeret}},
  \bibinfo{author}{\bibfnamefont{A.~F.} \bibnamefont{Volkov}},
  \bibnamefont{and} \bibinfo{author}{\bibfnamefont{K.~B.}
  \bibnamefont{Efetov}}, \bibinfo{journal}{Rev. Mod. Phys.}
  \textbf{\bibinfo{volume}{77}}, \bibinfo{pages}{1321} (\bibinfo{year}{2005});
%\bibitem[{\citenamefont{Izyumov et~al.}(2002)\citenamefont{Izyumov, Proshin,
%  and Husainov}}]{izyumov}
\bibinfo{author}{\bibfnamefont{Y.~A.} \bibnamefont{Izyumov}},
  \bibinfo{author}{\bibfnamefont{Y.~N.} \bibnamefont{Proshin}},
  \bibnamefont{and} \bibinfo{author}{\bibfnamefont{M.~G.}
  \bibnamefont{Husainov}},
\bibinfo{journal}{Sov. Phys.--Uspekhi} \textbf{\bibinfo{volume}{45}}, \bibinfo{pages}{109} (\bibinfo{year}{2002});
%\bibitem[{\citenamefont{Velez et~al.}(2008)\citenamefont{Velez, Martin,
%  Villegas, Hoffmann, Gonzalez, Vicent, and Schuller}}]{schrev}
\bibinfo{author}{\bibfnamefont{M.}~\bibnamefont{Velez}},
  \bibinfo{author}{\bibfnamefont{J.~I.} \bibnamefont{Martin}},
  \bibinfo{author}{\bibfnamefont{J.~E.} \bibnamefont{Villegas}},
  \bibinfo{author}{\bibfnamefont{A.}~\bibnamefont{Hoffmann}},
  \bibinfo{author}{\bibfnamefont{E.~M.} \bibnamefont{Gonzalez}},
  \bibinfo{author}{\bibfnamefont{J.~L.} \bibnamefont{Vicent}},
  \bibnamefont{and} \bibinfo{author}{\bibfnamefont{I.~K.}
  \bibnamefont{Schuller}},
\bibinfo{journal}{J. Magn. Magn. Mater.}
  \textbf{\bibinfo{volume}{320}}, \bibinfo{pages}{2547} (\bibinfo{year}{2008}).

\bibitem[{\citenamefont{Yang et~al.}(2004)\citenamefont{Yang, Lange, Volodin,
  Szymczak, and Moshchalkov}}]{vvmnatmat}
\bibinfo{author}{\bibfnamefont{Z.}~\bibnamefont{Yang}},
  \bibinfo{author}{\bibfnamefont{M.}~\bibnamefont{Lange}},
  \bibinfo{author}{\bibfnamefont{A.}~\bibnamefont{Volodin}},
  \bibinfo{author}{\bibfnamefont{R.}~\bibnamefont{Szymczak}}, \bibnamefont{and}
  \bibinfo{author}{\bibfnamefont{V.~V.} \bibnamefont{Moshchalkov}},
  \bibinfo{journal}{Nat. Mater.} \textbf{\bibinfo{volume}{3}},
  \bibinfo{pages}{793} (\bibinfo{year}{2004}).

\bibitem[{\citenamefont{Bulaevskii et~al.}(2000)\citenamefont{Bulaevskii,
  Chudnovsky, and Maley}}]{bulaevskiAPL}
\bibinfo{author}{\bibfnamefont{L.~N.} \bibnamefont{Bulaevskii}},
  \bibinfo{author}{\bibfnamefont{E.~M.} \bibnamefont{Chudnovsky}},
  \bibnamefont{and} \bibinfo{author}{\bibfnamefont{M.~P.} \bibnamefont{Maley}},
  \bibinfo{journal}{Appl. Phys. Lett.} \textbf{\bibinfo{volume}{76}},
  \bibinfo{pages}{2594} (\bibinfo{year}{2000}).

\bibitem[{\citenamefont{Gillijns et~al.}(2005)\citenamefont{Gillijns,
  Aladyshkin, Lange, Bael, and Moshchalkov}}]{gillijns}
\bibinfo{author}{\bibfnamefont{W.}~\bibnamefont{Gillijns}},
  \bibinfo{author}{\bibfnamefont{A.~Y.} \bibnamefont{Aladyshkin}},
  \bibinfo{author}{\bibfnamefont{M.}~\bibnamefont{Lange}},
  \bibinfo{author}{\bibfnamefont{M.~J.~V.} \bibnamefont{Bael}},
  \bibnamefont{and} \bibinfo{author}{\bibfnamefont{V.~V.}
  \bibnamefont{Moshchalkov}},
\bibinfo{journal}{Phys. Rev. Lett.} \textbf{\bibinfo{volume}{95}}, \bibinfo{pages}{227003}
  (\bibinfo{year}{2005}).

\bibitem[{\citenamefont{Stamopoulos et~al.}(2005)\citenamefont{Stamopoulos,
  Pissas, and Manios}}]{stamopoulos}
\bibinfo{author}{\bibfnamefont{D.}~\bibnamefont{Stamopoulos}},
  \bibinfo{author}{\bibfnamefont{M.}~\bibnamefont{Pissas}}, \bibnamefont{and}
  \bibinfo{author}{\bibfnamefont{E.}~\bibnamefont{Manios}},
  \bibinfo{journal}{Phys. Rev. B} \textbf{\bibinfo{volume}{71}},
  \bibinfo{pages}{014522} (\bibinfo{year}{2005}).

\bibitem[{\citenamefont{Milo\v{s}evi\'{c} and Peeters}(2005)}]{milosevic}
\bibinfo{author}{\bibfnamefont{M.~V.} \bibnamefont{Milo\v{s}evi\'{c}}}
  \bibnamefont{and} \bibinfo{author}{\bibfnamefont{F.~M.}
  \bibnamefont{Peeters}}, \bibinfo{journal}{Europhys. Lett.}
  \textbf{\bibinfo{volume}{70}}, \bibinfo{pages}{670} (\bibinfo{year}{2005}).

\bibitem[{\citenamefont{Vlasko-Vlasov et~al.}(2008)\citenamefont{Vlasko-Vlasov,
  Welp, Karapetrov, Novosad, Rosenmann, Iavarone, Belkin, and
  Kwok}}]{vlasko_hybrid}
\bibinfo{author}{\bibfnamefont{V.~V.} \bibnamefont{Vlasko-Vlasov}},
  \bibinfo{author}{\bibfnamefont{U.}~\bibnamefont{Welp}},
  \bibinfo{author}{\bibfnamefont{G.}~\bibnamefont{Karapetrov}},
  \bibinfo{author}{\bibfnamefont{V.}~\bibnamefont{Novosad}},
  \bibinfo{author}{\bibfnamefont{D.}~\bibnamefont{Rosenmann}},
  \bibinfo{author}{\bibfnamefont{M.}~\bibnamefont{Iavarone}},
  \bibinfo{author}{\bibfnamefont{A.}~\bibnamefont{Belkin}}, \bibnamefont{and}
  \bibinfo{author}{\bibfnamefont{W.~K.} \bibnamefont{Kwok}},
  \bibinfo{journal}{Phys. Rev. B} \textbf{\bibinfo{volume}{77}},
  \bibinfo{pages}{134518} (\bibinfo{year}{2008}).

\bibitem[{\citenamefont{Belkin et~al.}(2008{\natexlab{a}})\citenamefont{Belkin,
  Novosad, Iavarone, Fedor, Pearson, Petrean-Troncalli, and
  Karapetrov}}]{belkin_APL}
\bibinfo{author}{\bibfnamefont{A.}~\bibnamefont{Belkin}},
  \bibinfo{author}{\bibfnamefont{V.}~\bibnamefont{Novosad}},
  \bibinfo{author}{\bibfnamefont{M.}~\bibnamefont{Iavarone}},
  \bibinfo{author}{\bibfnamefont{J.}~\bibnamefont{Fedor}},
  \bibinfo{author}{\bibfnamefont{J.~E.} \bibnamefont{Pearson}},
  \bibinfo{author}{\bibfnamefont{A.}~\bibnamefont{Petrean-Troncalli}},
  \bibnamefont{and}
  \bibinfo{author}{\bibfnamefont{G.}~\bibnamefont{Karapetrov}},
  \bibinfo{journal}{Appl. Phys. Lett.} \textbf{\bibinfo{volume}{93}},
  \bibinfo{pages}{072510} (\bibinfo{year}{2008}{\natexlab{a}}).

\bibitem[{\citenamefont{Brongersma et~al.}(1993)\citenamefont{Brongersma,
  Verweij, Koeman, de~Groot, Griessen, and Ivlev}}]{ivlev}
\bibinfo{author}{\bibfnamefont{S.~H.} \bibnamefont{Brongersma}},
  \bibinfo{author}{\bibfnamefont{E.}~\bibnamefont{Verweij}},
  \bibinfo{author}{\bibfnamefont{N.~J.} \bibnamefont{Koeman}},
  \bibinfo{author}{\bibfnamefont{D.~G.} \bibnamefont{de~Groot}},
  \bibinfo{author}{\bibfnamefont{R.}~\bibnamefont{Griessen}}, \bibnamefont{and}
  \bibinfo{author}{\bibfnamefont{B.~I.} \bibnamefont{Ivlev}},
  \bibinfo{journal}{Phys. Rev. Lett.} \textbf{\bibinfo{volume}{71}},
  \bibinfo{pages}{2319} (\bibinfo{year}{1993}).

\bibitem[{\citenamefont{Guimpel et~al.}(1988)\citenamefont{Guimpel, Civale,
  de~la Cruz, Murduck, and Schuller}}]{guimpel}
\bibinfo{author}{\bibfnamefont{J.}~\bibnamefont{Guimpel}},
  \bibinfo{author}{\bibfnamefont{L.}~\bibnamefont{Civale}},
  \bibinfo{author}{\bibfnamefont{F.}~\bibnamefont{de~la Cruz}},
  \bibinfo{author}{\bibfnamefont{J.}~\bibnamefont{Murduck}}, \bibnamefont{and}
  \bibinfo{author}{\bibfnamefont{I.~K.} \bibnamefont{Schuller}},
  \bibinfo{journal}{Phys. Rev. B} \textbf{\bibinfo{volume}{38}},
  \bibinfo{pages}{2342} (\bibinfo{year}{1988}).

\bibitem[{\citenamefont{Andre et~al.}(2000)\citenamefont{Andre, Polichetti,
  Pastoriza, and Kes}}]{kes_channels}
\bibinfo{author}{\bibfnamefont{M.~O.} \bibnamefont{Andre}},
  \bibinfo{author}{\bibfnamefont{M.}~\bibnamefont{Polichetti}},
  \bibinfo{author}{\bibfnamefont{H.}~\bibnamefont{Pastoriza}},
  \bibnamefont{and} \bibinfo{author}{\bibfnamefont{P.~H.} \bibnamefont{Kes}},
  \bibinfo{journal}{Physica C} \textbf{\bibinfo{volume}{338}},
  \bibinfo{pages}{179} (\bibinfo{year}{2000}).

\bibitem[{\citenamefont{Kokubo et~al.}(2002)\citenamefont{Kokubo, Besseling,
  Vinokur, and Kes}}]{kokubo}
\bibinfo{author}{\bibfnamefont{N.}~\bibnamefont{Kokubo}},
  \bibinfo{author}{\bibfnamefont{R.}~\bibnamefont{Besseling}},
  \bibinfo{author}{\bibfnamefont{V.~M.} \bibnamefont{Vinokur}},
  \bibnamefont{and} \bibinfo{author}{\bibfnamefont{P.~H.} \bibnamefont{Kes}},
  \bibinfo{journal}{Phys. Rev. Lett.} \textbf{\bibinfo{volume}{88}},
  \bibinfo{pages}{247004} (\bibinfo{year}{2002}).

\bibitem[{\citenamefont{Stan et~al.}(2004)\citenamefont{Stan, Field, and
  Martinis}}]{field}
\bibinfo{author}{\bibfnamefont{G.}~\bibnamefont{Stan}},
  \bibinfo{author}{\bibfnamefont{S.~B.} \bibnamefont{Field}}, \bibnamefont{and}
  \bibinfo{author}{\bibfnamefont{J.~M.} \bibnamefont{Martinis}},
  \bibinfo{journal}{Phys. Rev. Lett.} \textbf{\bibinfo{volume}{92}},
  \bibinfo{pages}{097003} (\bibinfo{year}{2004}).

\bibitem[{\citenamefont{Grigorieva et~al.}(2004)\citenamefont{Grigorieva, Geim,
  Dubonos, Novoselov, Vodolazov, Peeters, Kes, and Hesselberth}}]{grigorieva}
\bibinfo{author}{\bibfnamefont{I.~V.} \bibnamefont{Grigorieva}},
  \bibinfo{author}{\bibfnamefont{A.~K.} \bibnamefont{Geim}},
  \bibinfo{author}{\bibfnamefont{S.~V.} \bibnamefont{Dubonos}},
  \bibinfo{author}{\bibfnamefont{K.~S.} \bibnamefont{Novoselov}},
  \bibinfo{author}{\bibfnamefont{D.~Y.} \bibnamefont{Vodolazov}},
  \bibinfo{author}{\bibfnamefont{F.~M.} \bibnamefont{Peeters}},
  \bibinfo{author}{\bibfnamefont{P.~H.} \bibnamefont{Kes}}, \bibnamefont{and}
  \bibinfo{author}{\bibfnamefont{M.}~\bibnamefont{Hesselberth}},
  \bibinfo{journal}{Phys. Rev. Lett.} \textbf{\bibinfo{volume}{92}},
  \bibinfo{pages}{237001} (\bibinfo{year}{2004}).

\bibitem[{\citenamefont{Karapetrov et~al.}(2006)\citenamefont{Karapetrov,
  Fedor, Iavarone, Rosenmann, and Kwok}}]{karapetrov_PRL}
\bibinfo{author}{\bibfnamefont{G.}~\bibnamefont{Karapetrov}},
  \bibinfo{author}{\bibfnamefont{J.}~\bibnamefont{Fedor}},
  \bibinfo{author}{\bibfnamefont{M.}~\bibnamefont{Iavarone}},
  \bibinfo{author}{\bibfnamefont{D.}~\bibnamefont{Rosenmann}},
  \bibnamefont{and} \bibinfo{author}{\bibfnamefont{W.~K.} \bibnamefont{Kwok}},
  \bibinfo{journal}{Phys. Rev. Lett.} \textbf{\bibinfo{volume}{95}},
  \bibinfo{pages}{167002} (\bibinfo{year}{2006}).

%\bibitem[{\citenamefont{Hess et~al.}(1990)\citenamefont{Hess, Robinson, and
%  Waszczak}}]{hess}
%\bibinfo{author}{\bibfnamefont{H.~F.} \bibnamefont{Hess}},
%  \bibinfo{author}{\bibfnamefont{R.~B.} \bibnamefont{Robinson}},
%  \bibnamefont{and} \bibinfo{author}{\bibfnamefont{J.~V.}
%  \bibnamefont{Waszczak}}, \bibinfo{journal}{Phys. Rev. Lett.}
%  \textbf{\bibinfo{volume}{64}}, \bibinfo{pages}{2711} (\bibinfo{year}{1990}).

\bibitem[{\citenamefont{Karapetrov et~al.}(2005)\citenamefont{Karapetrov,
  Fedor, Iavarone, Marshall, and Divan}}]{karapetrov_APL05}
\bibinfo{author}{\bibfnamefont{G.}~\bibnamefont{Karapetrov}},
  \bibinfo{author}{\bibfnamefont{J.}~\bibnamefont{Fedor}},
  \bibinfo{author}{\bibfnamefont{M.}~\bibnamefont{Iavarone}},
  \bibinfo{author}{\bibfnamefont{M.~T.} \bibnamefont{Marshall}},
  \bibnamefont{and} \bibinfo{author}{\bibfnamefont{R.}~\bibnamefont{Divan}},
  \bibinfo{journal}{Appl. Phys. Lett.} \textbf{\bibinfo{volume}{87}},
  \bibinfo{pages}{162515} (\bibinfo{year}{2005}).

\bibitem[{\citenamefont{Belkin}(2008{\natexlab{b}})\citenamefont{Belkin,
  Novosad, Iavarone, Fedor, Pearson, and Karapetrov}}]{belkin_PRB}
\bibinfo{author}{\bibfnamefont{A.}~\bibnamefont{Belkin}},
  \bibinfo{author}{\bibfnamefont{V.}~\bibnamefont{Novosad}},
  \bibinfo{author}{\bibfnamefont{M.}~\bibnamefont{Iavarone}},
  \bibinfo{author}{\bibfnamefont{J.}~\bibnamefont{Fedor}},
  \bibinfo{author}{\bibfnamefont{J.~E.} \bibnamefont{Pearson}},
  \bibnamefont{and}
  \bibinfo{author}{\bibfnamefont{G.}~\bibnamefont{Karapetrov}},
  \bibinfo{journal}{Phys. Rev. B} \textbf{\bibinfo{volume}{77}},
  \bibinfo{pages}{180506} (\bibinfo{year}{2008}{\natexlab{b}}).

\bibitem[{\citenamefont{Novoselov}(2005)\citenamefont{Novoselov, Jiang,
  Schedin, Booth, Khotkevich, Morozov, and Geim}}]{novoselov_PNAS}
 \bibinfo{author}{\bibfnamefont{K.~S.} \bibnamefont{Novoselov}},
  \bibinfo{author}{\bibfnamefont{D.}~\bibnamefont{Jiang}},
  \bibinfo{author}{\bibfnamefont{F.}~\bibnamefont{Schedin}},
  \bibinfo{author}{\bibfnamefont{T.~J.} \bibnamefont{Booth}},
  \bibinfo{author}{\bibfnamefont{V.~V.} \bibnamefont{Khotkevich}},
  \bibinfo{author}{\bibfnamefont{S.~V.} \bibnamefont{Morozov}},
  \bibnamefont{and} \bibinfo{author}{\bibfnamefont{A.~K.} \bibnamefont{Geim}},
  \bibinfo{journal}{Proc. Nat. Acad. Sci. USA} \textbf{\bibinfo{volume}{102}},
  \bibinfo{pages}{10451} (\bibinfo{year}{2005}).

\bibitem[{\citenamefont{Saito}(1964)\citenamefont{Saito, Fujiwara, and
  Sugita}}]{saito}
\bibinfo{author}{\bibfnamefont{N.}~\bibnamefont{Saito}},
  \bibinfo{author}{\bibfnamefont{H.}~\bibnamefont{Fujiwara}}, \bibnamefont{and}
  \bibinfo{author}{\bibfnamefont{Y.}~\bibnamefont{Sugita}},
  \bibinfo{journal}{J. Phys. Soc. Japan} \textbf{\bibinfo{volume}{19}},
  \bibinfo{pages}{1116} (\bibinfo{year}{1964}).

\bibitem[{\citenamefont{Spain}(1963)}]{spain}
\bibinfo{author}{\bibfnamefont{R.}~\bibnamefont{Spain}},
  \bibinfo{journal}{Appl. Phys. Lett.} \textbf{\bibinfo{volume}{3}},
  \bibinfo{pages}{208} (\bibinfo{year}{1963}).

\bibitem[{\citenamefont{Szymczak}(1968)}]{szym}
\bibinfo{author}{\bibfnamefont{R.}~\bibnamefont{Szymczak}},
  \bibinfo{journal}{J. Appl. Phys.} \textbf{\bibinfo{volume}{39}},
  \bibinfo{pages}{875} (\bibinfo{year}{1968}).

\bibitem[{\citenamefont{Milo\v{s}evi\'{c} and Peeters}(2004)}]{miloPRL}
\bibinfo{author}{\bibfnamefont{M.~V.} \bibnamefont{Milo\v{s}evi\'{c}}}
  \bibnamefont{and} \bibinfo{author}{\bibfnamefont{F.~M.}
  \bibnamefont{Peeters}}, \bibinfo{journal}{Phys. Rev. Lett.}
  \textbf{\bibinfo{volume}{93}}, \bibinfo{pages}{267006}
  (\bibinfo{year}{2004}).

\bibitem[{\citenamefont{Reichhardt et~al.}(2006)\citenamefont{Reichhardt,
  Lib{\"a}l, and Reichhardt}}]{reichhardt_PRB73}
\bibinfo{author}{\bibfnamefont{C.~J.~O.} \bibnamefont{Reichhardt}},
  \bibinfo{author}{\bibfnamefont{A.}~\bibnamefont{Lib{\"a}l}},
  \bibnamefont{and}
  \bibinfo{author}{\bibfnamefont{C.}~\bibnamefont{Reichhardt}},
  \bibinfo{journal}{Phys. Rev. B} \textbf{\bibinfo{volume}{73}},
  \bibinfo{pages}{184519} (\bibinfo{year}{2006}).

\bibitem[{\citenamefont{Milo\v{s}evi\'{c} and Peeters}(2003)}]{miloPRB}
\bibinfo{author}{\bibfnamefont{M.~V.} \bibnamefont{Milo\v{s}evi\'{c}}}
  \bibnamefont{and} \bibinfo{author}{\bibfnamefont{F.}~\bibnamefont{Peeters}},
  \bibinfo{journal}{Phys. Rev. B} \textbf{\bibinfo{volume}{68}},
  \bibinfo{pages}{094510} (\bibinfo{year}{2003}).

\bibitem[{\citenamefont{Liu et~al.}(2003)\citenamefont{Liu, Avinash, and
  Goree}}]{liu}
\bibinfo{author}{\bibfnamefont{B.}~\bibnamefont{Liu}},
  \bibinfo{author}{\bibfnamefont{K.}~\bibnamefont{Avinash}}, \bibnamefont{and}
  \bibinfo{author}{\bibfnamefont{J.}~\bibnamefont{Goree}},
  \bibinfo{journal}{Phys. Rev. Lett.} \textbf{\bibinfo{volume}{91}},
  \bibinfo{pages}{255003} (\bibinfo{year}{2003}).

\bibitem[{\citenamefont{Piacente et~al.}(2004)\citenamefont{Piacente,
  Schweigert, Betouras, and Peeters}}]{piac}
\bibinfo{author}{\bibfnamefont{G.}~\bibnamefont{Piacente}},
  \bibinfo{author}{\bibfnamefont{I.~V.} \bibnamefont{Schweigert}},
  \bibinfo{author}{\bibfnamefont{J.~J.} \bibnamefont{Betouras}},
  \bibnamefont{and} \bibinfo{author}{\bibfnamefont{F.~M.}
  \bibnamefont{Peeters}}, \bibinfo{journal}{Phys. Rev. B}
  \textbf{\bibinfo{volume}{69}}, \bibinfo{pages}{045324}
  (\bibinfo{year}{2004}).

\bibitem[{\citenamefont{Haghgooie and Doyle}(2004)}]{hagh}
\bibinfo{author}{\bibfnamefont{R.}~\bibnamefont{Haghgooie}} \bibnamefont{and}
  \bibinfo{author}{\bibfnamefont{P.~S.} \bibnamefont{Doyle}},
  \bibinfo{journal}{Phys. Rev. E} \textbf{\bibinfo{volume}{70}},
  \bibinfo{pages}{061408} (\bibinfo{year}{2004}).



\end{thebibliography}

\end{document}